\RequirePackage{fix-cm}
\documentclass[smallextended]{svjour3}       
\smartqed  
\usepackage{graphics,epstopdf}
\usepackage{graphicx}
\usepackage[colorlinks=true, citecolor=blue, urlcolor=blue ]{hyperref}
\usepackage{float}

\begin{document}
\title{Classical communication and non-classical fidelity of quantum teleportation}
\author{Manik Banik         \and
        Md. Rajjak Gazi 
}
\institute{Manik Banik \at
              Physics and Applied Mathematics Unit, Indian Statistical Institute, 203 B.T. Road, Kolkata-700108, India\\
              \email{manik11ju@gmail.com}           
           \and
           Md. Rajjak Gazi\at
              Physics and Applied Mathematics Unit, Indian Statistical Institute, 203 B.T. Road, Kolkata-700108, India\\
                            \email{rajjakgazimath@gmail.com}           
}

\date{Received: date / Accepted: date}



\maketitle

\begin{abstract}
In quantum teleportation, the role of entanglement has been much discussed. It is known that entanglement is necessary for achieving non-classical teleportation fidelity. Here we focus on the amount of classical communication that is necessary to obtain non-classical fidelity in teleportation. We quantify the amount of classical communication that is sufficient for achieving non-classical fidelity for two independent $1$-bit and single $2$-bits noisy classical channels.  It is shown that on average $0.208$ bits of classical communication is sufficient to get non-classical fidelity. We also find the necessary amount of classical communication in case of isotropic transformation. Finally we study how the amount of sufficient classical communication increases with weakening of entanglement used in the teleportation process. 

\end{abstract}

\section{Introduction}
Quantum information theory, the generalization of classical information theory governed by the rules of quantum mechanics, has been a subject of much recent interest. It is now known that various interesting communication schemes, like teleportation \cite{bennett,rie,furu} and dense coding \cite{bennett1,mattle,braunstein}, are possible when the laws of quantum mechanics are invoked in information processing. It was customary to ignore the classical communication cost in different quantum information processes like entanglement distillation \cite{bennett2,bennett4,pan}. The motivation was that classical communication is free whereas entanglement is expensive. However, as emphasized in \cite{lo}, in dense coding process classical communication cost is of primary interest and one can not ignore it. On the other hand in quantum teleportation protocol classical communication is one of the required resource other than entanglement. An unknown qubit can be teleportated perfectly by using a maximally entangled state and $2$ bits of classical communication. Naturally, accuracy of the teleportation process decreases if noise is introduced in the quantum as well as classical channel. A lot of works has been devoted to study teleportation process in presence of noisy quantum channel \cite{popescu,Oh,Horodecki,Banaszek}, but very little is discussed about the efficiency of teleportation in case where classical channel is noisy. So it is also important to study quantum teleportation in presence of noisy classical channel.

Suppose there are two spatially separated fellows, say Alice and Bob. Alice is holding a spin-$\frac{1}{2}$ system in state $|\phi\rangle$, where the state of the system $|\phi\rangle$ is unknown to her. Now challenge for them is to prepare the state $|\phi\rangle$ at Bob's location. Bennett \emph{et.al.} have shown that this seemingly impossible task can be done in quantum mechanical framework, which is known as quantum teleportation \cite{bennett}. To be successful in this challenge, Alice and Bob must share two different types of resources namely entanglement and classical communication. In general they will not succeed in perfectly reproducing the state $|\phi\rangle$ at Bob's location, but will prepare some other state, denoted by $|\phi'\rangle$ if it is pure, or by $\chi$ (density matrix) if it is mixed. The measure of success - the ``score" obtained can be quantified by $|\langle\phi|\phi'\rangle|^2$ or equivalently by $\mbox{Tr}(\chi|\phi\rangle\langle\rangle\phi|)$. Suppose now that this process is repeated many times, each time Alice gets an unknown particle uniformly distributed on the unit sphere. The final score of the process - the average of the scores obtained in the different runs - is called the fidelity of the transmission \cite{popescu}. If Alice and Bob cannot communicate at all, Bob could try to guess the state, according to some particular scheme ( one of such scheme is that Bob will always prepare a particle in some state). The fidelity of any such guess scheme will be always $\frac{1}{2}$ \cite{popescu,masar}. 

If Alice and Bob use no entanglement but classical communications then they can raise the fidelity value beyond $\frac{1}{2}$. Alice measures her state and sends the result to Bob, who makes his best guess of the state based on this information. The success of sending the state $|\phi\rangle$ is calculated as $F_{cl}(|\phi\rangle)=\sum^n_{i=1}p(i||\phi\rangle)|\langle\phi|\alpha_i\rangle|^2$, where $p(i||\phi\rangle)=\langle\phi|A_i|\phi\rangle$ is the probability of Alice obtaining the result corresponding to the positive operator $A_i$ out of $n$ possible outcomes of the positive operator valued measure (POVM) $\{A_i\}$ \cite{Kraus,Peres}, where $\sum^n_{i=1}A_i=\textbf{1}_2$, with $\textbf{1}_2$ denoting the identity operator; and $|\alpha_i\rangle$ be the Bob's guess, given outcome $i$. When the input state is completely unknown, the average of the success over a uniform distribution of all states on the Bloch sphere is taken to obtain the fidelity of the teleportation process \cite{hen,horo}. It has been shown  that the fidelity in this case becomes $F_{cl}=\int\sum^n_{i=1}p(i||\phi\rangle)|\langle\phi|\alpha_i\rangle|^2 d\Omega=\frac{2}{3}$ \cite{masar}, here $d\Omega$ denote the haar measure over the set of pure states on the Bloch sphere and the subindex `$cl$' denotes that only classical communication has been used as teleportation resource. This value we call \emph{classical fidelity} of teleportation process. In \cite{popescu}, Popescu showed that if Alice performed a two outcomes projective measurement in any basis (say $\sigma_z$) and communicated her measurement result to Bob, who prepared the particle either in $\sigma_z$-up or $\sigma_z$-down state depending on Alice's measurement result, then optimal classical fidelity (i.e. $\frac{2}{3}$) can be achieved. If in any teleportation scheme the value of the teleportation fidelity becomes greater than $\frac{2}{3}$, then one can conclude that  some quantum entanglement must have been used in the process. When $2$ bits of classical communication and maximally entangled state (e.g. singlet) are used in teleportation protocol the value of teleportation fidelity becomes unity, which reflects complete accuracy of the process. If the degree of entanglement used in the teleportation process is weakened in the form of Werner state $W_{\alpha} = \alpha|\psi^{singlet}\rangle\langle\psi^{singlet}|+(1-\alpha)\frac{\textbf{1}_2}{2}\otimes\frac{\textbf{1}_2}{2}$, then teleportation fidelity  takes the value $F=\frac{1+\alpha}{2}$ \cite{popescu}. The parameter $(1-\alpha)$ in the Werner state $W_{\alpha}$ quantifies the amounts of \emph{white noise} (i.e. $\frac{\textbf{1}_2}{2}\otimes \frac{\textbf{1}_2}{2}$) present in the state $W_{\alpha}$ along with the singlet state. It is known that the state $W_{\alpha}$ is entangled for $\alpha>\frac{1}{3}$ and therefore  from the expression of $F (=\frac{1+\alpha}{2})$ one can infer that to achieve the fidelity greater than the classical value the Werner state has to be necessarily entangled. Interestingly it has been shown that any two-qubit entangled state is good for teleportation \cite{Verstraete}. In this paper we pose the following question: how much  classical communication is necessary to achieve \emph{non-classical fidelity} in quantum teleportation process when there is no restriction on entanglement? Interestingly we have found that:

(a) If two independent $1$-bit noisy classical channels are used then on average $(0.255+\epsilon)$ bits of classical communication (with $\epsilon>0$) is sufficient to make the teleportation fidelity non-classical.

(b) In case of $2$-bits noisy classical channel, on average $(0.208+\epsilon)$ bits of classical communication (with $\epsilon>0$) is sufficient to make the teleportation fidelity non-classical. We also argue that $(0.208+\epsilon)$ bits of classical communication is necessary if the quantum state transforms isotropically under teleportation process.

(c) We have shown how the amount of classical communication increases when the entanglement, used in the teleportation process, is weakened.   

The paper is organized in the following way. First, in section-\ref{single} and section-\ref{double} we study teleportation process with $1$-bit noisy classical channel and $2$-bits noisy classical channel respectively. Then in section-\ref{necessery} we discuss about necessary amount of classical communication required to get non classical fidelity when state transforms isotropically. In section-\ref{werner} we find how the amount of classical communication increases to get non classical fidelity when entangled channel is weakened in Werner form.  

\section{Teleportation with $1$-bit noisy classical channel}\label{single}
Suppose Alice has to teleport an unknown qubit state $|\phi_1\rangle=a|0_1\rangle+b|1_1\rangle$ to Bob, $|0\rangle$ and $|1\rangle$ respectively denote the up and down eigenstate of spin observable along $z$-direction and the subindex is used to denote the particle. They have a shared singlet pair $|\psi^-_{23}\rangle =\frac{|0_21_3\rangle -|1_20_3\rangle}{\sqrt{2}}$, i.e. they are allowed to share $1$-ebit entanglement as teleportation resource, and they have two independent $1$-bit noisy channels $C_\eta$ and $C_\delta$. The $1$-bit noisy classical channel $C_z$ $( z \equiv \eta,\delta)$ is characterized by the conditional probability :-
\begin{eqnarray}
p(a|a) = z\nonumber\\
p(\bar{a}|a) = 1-z
\end{eqnarray}
where $a\in\{0,1\}$ is Alice's input bit; $\bar{a}$ denote complement bit of $a$. The value of $z$ lies between $\frac{1}{2}$ and $1$. The amount of classical communication is quantified as mutual information \cite{thomas,nielsen,wilde} between Alice's input and Bob's output, which is in this case :-
\begin{equation}
C_z:=1-H(z)
\end{equation}
where $H(z)=-z\log_2z-(1-z)\log_2(1-z)$ is the Shannon entropy \cite{thomas,nielsen}.

The teleportation protocol goes as standard teleportation protocol \cite{bennett} : Alice first performs measurement on the joint system of unknown qubit (particle $1$) and her part (particle $2$) of the entangled pair in Bell basis $(|\psi^\pm\rangle,|\varphi^\pm\rangle)$, Where
\begin{equation}
|\psi^\pm\rangle =\frac{|01\rangle\pm|10\rangle}{\sqrt{2}}~~~;~~~~ |\varphi^\pm\rangle=\frac{|00\rangle\pm|11\rangle}{\sqrt{2}}
\end{equation}
After the measurement Alice sends the results to Bob by using the two noisy classical channels. Bob, after receiving the classical bit, performs the unitary operation on his part (particle $3$) of the entangled pair as required for perfect teleportation.
This results the following density matrix for Bob's qubit;
\begin{eqnarray}\label{singl}
\varrho'=\eta\delta\varrho+\eta(1-\delta)\sigma_z\varrho\sigma_z+\delta(1-\eta)\sigma_x\varrho\sigma_x,\nonumber\\
~~~~~~~~~~~~~~~~~+(1-\eta)(1-\delta)\sigma_y\varrho\sigma_y
\end{eqnarray}
where $\varrho=|\phi_1\rangle\langle\phi_1|$. Measure of success is given by the quantity $\mbox{Tr}{[\varrho\varrho']}$. Average of this quantity over all possible pure states gives us the fidelity of the teleportation process \cite{popescu}. For a qubit system the density matrix $\varrho$ for a pure state can be written as $\frac{1}{2}(1+\vec n.\vec{\sigma})$ where $\vec n \equiv(n_x,n_y,n_z)\equiv(\sin\theta \cos\phi,\sin\theta \sin\phi,\cos\theta)$ is unit vector on Poincaré sphere and $\vec{\sigma}\equiv(\sigma_x,\sigma_y,\sigma_z)$ where $\sigma_x,\sigma_y,\sigma_z$ are the Pauli operator. Therefore the fidelity of the teleportation process is the following;
\begin{eqnarray}
F  &=&\langle \mbox{Tr}[\varrho(\eta\delta\varrho+\eta(1-\delta)\sigma_z\varrho\sigma_z+
\delta(1-\eta)\sigma_x\varrho\sigma_x\nonumber\\
&&+(1-\eta)(1-\delta)\sigma_y\varrho\sigma_y)]\rangle\nonumber\\
&=&\int \mbox{d}\varrho\mbox{Tr}[\varrho(\eta\delta\varrho+\eta(1-\delta)\sigma_z\varrho\sigma_z+
\delta(1-\eta)\sigma_x\varrho\sigma_x\nonumber\\
&&+(1-\eta)(1-\delta)\sigma_y\varrho\sigma_y)]\nonumber\\
&=&\int\mbox{d}\vec{n}[\eta\delta+\eta(1-\delta)n^2_z+\delta(1-\eta)n^2_x\nonumber\\
&&+(1-\eta)(1-\delta)n^2_y]\nonumber\\
&=&\frac{1}{4\pi}\int\sin\theta\mbox{d}\theta\mbox{d}\phi[\eta\delta+\eta(1-\delta)\cos^2\theta\nonumber\\
&&+\delta(1-\eta)\sin^2\theta\cos^2\phi+(1-\eta)(1-\delta)\sin^2\theta\sin^2\phi]\nonumber\\
&=&\eta\delta+\frac{1}{3}[\eta(1-\delta)+
\delta(1-\eta)+(1-\eta)(1-\delta)]\nonumber\\
&=& \frac{1+2\eta\delta}{3}
\end{eqnarray}

In teleportation process described above, the total classical communication used amounts to sum of the capacity (mutual information) of the two independent noisy classical channel i.e. $[2-H(\eta)-H(\delta)]$ claccical bit (cbit). The fidelity of the teleportation will go beyond the classical limit $(i.e. F >F_{cl}=\frac{2}{3})$ if we have $\eta\delta>\frac{1}{2}$. With the restriction $\eta\delta>\frac{1}{2}$ on the parameter $\eta$ and $\delta$, the minimum of the classical cost is found when $\eta = \delta = \frac{1}{\sqrt{2}}$. The optimal communication has been found to be $[2-H(\eta)-H(\delta)]\approx 0.255$ cbit. Thus we can say that on average $(0.255+\epsilon)$ bits of classical communication with $\epsilon>0$, is sufficient for achieving non-classical fidelity in the teleportation process when two independent noisy classical channels have been used.

\section{Teleportation with $2$-bits noisy classical channel}\label{double}
Let Alice and Bob are sharing a $2$-bits noisy classical channel which is characterized by following conditional probability :-
\begin{eqnarray}
P(a b|a b) = p_1\nonumber\\
P(\bar{a} b|a b) = p_2\nonumber\\
P(a \bar{b}|a b) = p_3\nonumber\\
P(\bar{a} \bar{b}|a b) = p_4
\end{eqnarray}
where $ab$ etc. are Alice's inputs and take values from the set $\{00,01,10,11\}$ and $p_i\geq 0$ for $i=1,2,3,4$ and $\Sigma_ip_i=1$.
The amount of classical communication is quantified as mutual information between Alice's input variables $X$ and Bob's outputs variables $Y$
\begin{eqnarray}
I(X:Y)&=& H(X)+H(Y)-H(X,Y)\nonumber\\
&=& 2+\Sigma^4_{i=1}~p_i \log_2 (p_i)
\end{eqnarray}
From the definition it is clear that at the two extreme cases, (i) one of $p_i=1$ and (ii) all $p_i$'s are equal, we have amount of classical communication $2$ bits and $0$ bit respectively.

If Alice and Bob follows the same teleportaion protocol as before, using the $2$-bits noisy classical channel, the resulting density matrix for Bob's qubit is as follows;
\begin{equation}\label{doubl}
\varrho' = p_1\varrho+p_2\sigma_x\varrho\sigma_x+p_3\sigma_z\varrho\sigma_z+p_4\sigma_y\varrho\sigma_y
\end{equation}
The teleportation fidelity in this case becomes,
\begin{eqnarray}
F &=& \langle \mbox{Tr}[\varrho(p_1\varrho+p_2\sigma_x\varrho\sigma_x+p_3\sigma_z\varrho\sigma_z+
p_4\sigma_y\varrho\sigma_y)]\rangle\nonumber\\
&=& p_1+\frac{1}{3}(p_2+p_3+p_4)\nonumber\\
&=& \frac{1+2p_1}{3}
\end{eqnarray}
The classical communication used in this process amounts to: 
\begin{equation}
C = I(X:Y) = 2+\Sigma^4_{i=1}p_i \log_2 (p_i)
\end{equation}
From Eqn.$(9)$ it is clear that we will get non classical fidelity ( i.e. $F>F_{cl}$) if we have $p_1>\frac{1}{2}$. With this restriction on $p_1$ we want to find the minimum amount of classical communication C. From the expression of C it can be shown that the  minimum will be achieved when $p_2 = p_3 = p_4$. Thus we have :-
\begin{equation}
C = 2+p_1 \log_2 p_1+(1-p_1)\log_2(\frac{1-p_1}{3})
\end{equation}
\begin{figure}[h!]
\centering
\includegraphics[height=5cm,width=8cm]{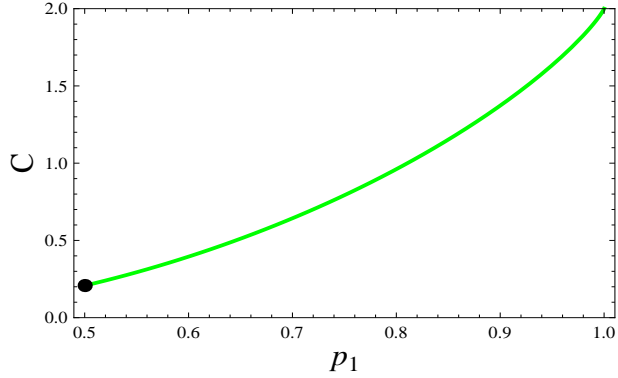}
\caption{The parameter $p_1$ is plotted along $x$ axis and along $y$ axis classical communication $C$ is plotted. Clearly $C$ is monotonic increasing function within the range $\frac{1}{2}\le p_1\le 1$. At $p_1=\frac{1}{2}$ we have $C\cong 0.208$. In the plot black dot indicates this point.}\label{fig1}
\end{figure}
In the range $\frac{1}{2}\leq p_1\leq 1$, C is a monotonic increasing function of $p_1$ and the value of C at $p_1=\frac{1}{2}$ is $C_{\frac{1}{2}}\approx 0.208$. Therefore we can say that on average $(0.208+\epsilon)$ bits of classical communication ($\epsilon>0$) is sufficient to get non classical teleportation fidelity. It is interesting to note that we are getting the optimal value of classical cost when $p_2 = p_3 = p_4$ and in this case the final density matrix is given by $\frac{1}{2}[1+(2f-1) \vec n.\sigma]$, where $f= \frac{2p_1+1}{3}$; which means that the optimality is achieved when the state transform isotropically.

\section{Necessary communication when state transforms isotropically}\label{necessery}
Now we will show that $0.208$ bits of classical communication is necessary to obtain non classical fidelity of the teleportation process which transform the state isotropically. To prove this, we assume there is a teleportation process which transform the state isotropically but at the same time attain non-classical fidelity by using $(0.208 - \epsilon)$ bits of classical communications. Let Alice and Bob share a singlet state and Alice in order to send two bits of classical information, applies the unitary transformation accordingly as is done in the standard dense coding protocol \cite{bennett2}. Then Alice teleports the qubit to Bob by the teleportation channel with non-classical fidelity which use $(0.208 - \epsilon)$ bits of classical communications and transform the state isotropically.

Depending on the Alice's applied unitary transformation, the four possible two-qubit density matrices on Bob's side are as follows:-
\begin{equation}\label{ensem}
\rho_i =\frac{(1-p_1)}{3}\textbf{1}_4+\frac{(4p_1-1)}{3}(\sigma_i\otimes\textbf{1}_2)|\psi^-\rangle\langle\psi^-|(\sigma_i\otimes\textbf{1}_2)
\end{equation}
where $\textbf{1}_4$ ($\textbf{1}_2$) be the $4\times4$ ($2\times2$) identity matrix; $i=0,x,y,z$ with $\sigma_0 = \textbf{1}_2$. All the density matrices $\rho_i$ occur with equal probability. The amount of classical bit that can be extracted \cite{nielsen,wilde} from an ensemble of density matrix $\{p_i,\pi_i\}$ ($\pi_i$'s are density matrices and $p_i$ be a valid probability distribution i.e. $p_i\ge 0$ and $\sum_ip_i=1$) is bounded by the Holevo quantity $\chi$ \cite{holevo} which is given by:
\begin{equation}
\chi:= S(\pi)-\sum_ip_iS(\pi_i),
\end{equation}
where $\pi=\sum_ip_i\Pi_i$ and $S(*)$ denote the Von Neumann entropy \cite{nielsen}. For the ensemble of the density matrices of Eqn.-(\ref{ensem}) the Holevo quantity becomes: 
\begin{equation}
\chi = 2+p_1 \log_2 p_1+(1-p_1)\log_2(\frac{1-p_1}{3})
\end{equation}
and $\chi$ is $(0.208 + \epsilon)$ for non-classical fidelity.  
Interestingly, this bound can be achieved by making measurement on  two-qubit in the Bell basis and Bob can extract classical information as done in super dense coding. But this implies that by communicating $(0.208 - \epsilon)$ bits of classical information, one can generate $0.208 $ bits, which violates causality principle. This proves that $(0.208 + \epsilon)$ bits of classical communication are necessary for teleportation which is isotropic and achieves quantum fidelity.
\section{Teleportation with Werner channel supported by noisy classical channel}\label{werner}
Let Alice and Bob share Werner state $W_{\alpha} =\alpha |\psi^-\rangle\langle\psi^-|+(1-\alpha)\frac{\textbf{1}_2}{2}\otimes\frac{\textbf{1}_2}{2}$ as resource for teleportation. And as classical supporting channel if they share $2$-bits noisy channel then the teleportation fidelity becomes: 
\begin{eqnarray}
F&=&\alpha\frac{1+2p_1}{3}+(1-\alpha)\frac{1}{2}\nonumber\\
&=&\frac{3-\alpha+4\alpha p_1}{6}
\end{eqnarray}
From the expression $F$ it is clear that for a given Werner state non classical fidelity will be achieved if $p_1\ge\frac{1+\alpha}{4\alpha}$. Therefore the amount of sufficient communication for achieving non classical fidelity when Werner state supported by $2$-bits noisy classical channel is given by $C(\alpha) = 2+\frac{1+\alpha}{4\alpha} \log_2(\frac{1+\alpha}{4\alpha})+\frac{3\alpha-1}{4\alpha}\log_2(\frac{3\alpha-1}{12\alpha})$.  

If $1$-bits noisy classical channel is shared between Alice and Bob along with the Werner channel then teleportation fidelity becomes:
\begin{equation}
F=\frac{3-\alpha+4\alpha \eta\delta}{6}
\end{equation}
Clearly in this case condition for getting non classical fidelity is $\eta\delta\ge\frac{1+\alpha}{4\alpha}$. When $1$-bit noisy classical channel is used the sufficient classical communication cost becomes $C'(\alpha) = 2[1+\sqrt{\frac{1+\alpha}{4\alpha}} \log_2(\sqrt{\frac{1+\alpha}{4\alpha}})+(1-\sqrt{\frac{1+\alpha}{4\alpha}}) \log_2(1-\sqrt{\frac{1+\alpha}{4\alpha}})]$.
\begin{figure}[h!]
\centering
\includegraphics[height=5cm,width=8cm]{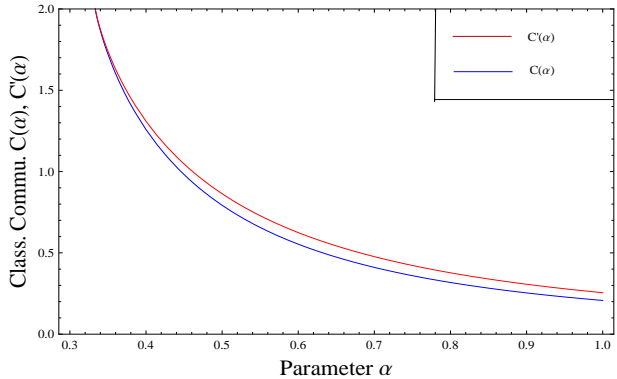}
\caption{This plot shows how sufficient classical communication changes with parameter $\alpha$ of the Werner state $W_{\alpha}$. Interestingly more the noise is introduced in entanglement, more the classical communication is needed to achieve \emph{non-classical fidelity}. From the plot it is clear that $2$-bits noisy classical channel is more useful than two independent $1$-bit noisy classical channel through the whole region of the parameter $\alpha$.}\label{fig2}
\end{figure}
For $\alpha=\frac{1}{3}$ Werner state becomes separable. Therefore using $W_{\frac{1}{3}}$ we can never achieve fidelity grater than $\frac{2}{3}$. From Fig-(\ref{fig2}) it is clear that if we follow the standard teleportation protocol using $W_{\frac{1}{3}}$ state supported with noisy classical channel then both for $2$-bits noisy classical channel and two independent $1$-bit noisy classical channels, the required clssical communication amounts to 2-cbits. Thus performing standard teleportation protocol is not a clever way when $W_{\frac{1}{3}}$ state is shared. In this case following the procedure of \cite{popescu} one can achieve $\frac{2}{3}$ fidelity with help of $1$-classical bit.  
 
\section{Discussion}
Our result shows that in quantum teleportation, there is a threshold amount of classical communication that is required for achieving non classical fidelity. The optimal amount of classical communication that is necessary for achieving non-classical fidelity in quantum teleportation still remains an open problem. We only provide sufficient amount in two cases. In case of isotropic state transformation, we find the necessary amount of classical communication to reach quantum fidelity. We also show that the sufficient amount of classical communications increase with weakening of the singlet state to Werner state. It remains interesting to study the necessary amount of classical communication for a general entangled state used as teleportation resources.  
\begin{acknowledgements}
We like to thank G.Kar for many simulating discussion and giving suggestions regarding the use of $2$-bits noisy classical channel. We also like to acknowledge S.Ghosh, S.Bandyopadhyay, S.Kunkri, A.Rai and S.Das for many helpful discussions. We would like to thank the anonymous referee for careful reading of the manuscript.
\end{acknowledgements}
\textbf{Note Added:} Recently  our work has been extended to qudits in \cite{Weinar} by Weinar \emph{et al}.

\end{document}